\documentstyle [aps,twocolumn,epsfig]{revtex}
\catcode`\@=11
\newcommand{\be}{\begin{equation}}
\newcommand{\ee}{\end{equation}}
\newcommand{\bea}{\begin{eqnarray}}
\newcommand{\eea}{\end{eqnarray}}

\newcommand{\up}{\uparrow}
\newcommand{\down}{\downarrow}

\newcommand{\ri}{\mbox{i}}

\begin {document}

\title{ Effect of Symmetry Breaking Perturbations in the One-Dimensional SU(4)
 Spin-Orbital Model}
\vspace{2cm}

\author{P. Azaria$^1$, E. Boulat$^1$, and   P. Lecheminant$^2$}

\vspace{0.5cm}

\address{$^1$ Laboratoire de Physique Th{\'e}orique des Liquides,
Universit{\'e} Pierre et Marie Curie, 4 Place Jussieu, 75252 Paris,
France\\
$^2$ Laboratoire de Physique Th{\'e}orique et Mod{\'e}lisation,
Universit{\'e} de Cergy-Pontoise, 5 Mail Gay-Lussac, Neuville sur Oise, 95301
Cergy-Pontoise Cedex, France} 
\vspace{3cm}

\address{\rm (Received: )}
\address{\mbox{ }}
\address{\parbox{14cm}{\rm \mbox{ }\mbox{ }
We study the effect of symmetry breaking perturbations in the one-dimensional 
SU(4) spin-orbital model. We allow the exchange in spin ($J_1$) and orbital
($J_2$) channel 
to be different and thus reduce the symmetry to SU(2) $\otimes$ SU(2).
A magnetic field $h$ along the $S^z$ direction is also applied. Using the 
formalism
developped by Azaria et al \cite{azaria} we extend their analysis of 
the isotropic $J_1 = J_2$, $h =0$ case and obtain the low-energy effective theory near the 
SU(4) 
point in the generic case $J_1 \neq J_2$, $h \neq 0$.
 In zero magnetic field, we retrieve the same qualitative low-energy 
physics than in the isotropic case. In particular, the isotropic massless 
behavior found on the line $J_1 = J_2 < K/4$ extends in a large
anisotropic region.  We discover however that the anisotropy plays its 
trick in allowing non trivial scaling behaviors of the physical quantities.
For example the mass gap $M$ has two different scaling behaviors depending
on the anisotropy. In addition, we show that in some regions,
the anisotropy is responsible for anomalous finite size effects and
may change qualitatively the shape of the computed critical line  in a finite
system. When a magnetic field is present the effect of the anisotropy is
striking. In addition to the usual commensurate-incommensurate 
phase transition that occurs in the spin sector of the theory, we find
that the field may induce a second transition of the KT type 
in the remaining degrees
of freedom to which it does $not$ couple directly. In this sector,
we find that the effective theory is that of an SO(4) Gross-Neveu model
with an $h$-dependent coupling that may change its sign as $h$ varies.}}
\address{\mbox{ }}
\address{\parbox{14cm}{\rm PACS No: 75.10.Jm, 75.40.Gb}}
\maketitle

\makeatletter
\global\@specialpagefalse 
\makeatother

\section{introduction}

In the very past years, there has been an intense interest devoted to 
one dimensional spin-orbital models\cite{kugel}. The main reason stems from the recent
discovery of the new quasi one dimensional spin gapped materials
Na$_2$Ti$_2$Sb$_2$O\cite{axtell} and NaV$_2$O$_5$\cite{isobe}.
It is believed that the unusual magnetic properties observed  in
these compounds should be explained by a simple two-band Hubbard  models
at $quarter$ filling. At this filling, and in the large Coulomb repulsion
the effective Hamiltonian simplifies greatly and is equivalent to
a model of two interacting spin  one-half Heisenberg models\cite{auerbach,pati}:
\bea 
{\cal H} = \sum_{i}   J_1 \; {\vec S}_{i} \cdot {\vec S}_{i+1}
&+& J_2 \; {\vec T}_{i} \cdot {\vec T}_{i+1} \nonumber \\
&+& K\left({\vec S}_{i} \cdot {\vec S}_{i+1}\right)
\left({\vec T}_{i} \cdot {\vec T}_{i+1}\right)
\label{hamilt}
\eea
where ${\vec S}_{i}$ and ${\vec T}_{i}$ are spin-1/2 operators
that represent spin and orbital degrees of freedom at each site $i$
and $J_{1(2)}$ and $K$ are positive.  
It is important to notice that (\ref{hamilt}) is not only SU(2) invariant
in $\vec S$ space but also in $\vec T$ space. For generic couplings 
$J_{1(2)}$ and $K$, the Hamiltonian  (\ref{hamilt}) is SU(2)$_s \otimes$ 
SU(2)$_t$ symmetric.
Two particular cases are  of interest. First, when
$J_{1}=J_{2}$  there is an additional $Z_2$ symmetry in the
exchange between ${\vec S}_{i}$ and ${\vec T}_{i}$. 
The other important
case is when $J_{1}=J_{2}= K/4$ in which case the Hamiltonian is SU(4)
invariant\cite{li}.  In fact, the Hamiltonian 
(\ref{hamilt}) is    a simplified version of the most general
case. Indeed, depending on the microscopic couplings, there can be
other terms that breaks  the SU(2) symmetry  in both spin and orbital 
sectors. The choice of studying (\ref{hamilt}) is dictated by the fact
that since it retains some symmetries  it is a simple starting point from 
which one can expect to gain some insight before tackling with
the more general case. In this respect, the model (\ref{hamilt}) describes the 
simplest physically relevant symmetry breaking pattern
SU(4) $\to$ SU(2)$_s \otimes$ SU(2)$_t$. 

The Hamiltonian (\ref{hamilt}) can  be interpreted in 
terms of a two-leg spin ladder coupled by a four spins 
interaction. 
Such an interaction can be generated either 
by phonons or, in the doped state, by conventional 
Coulomb repulsion between the holes\cite{nersesyan}. 
Some of the properties of (\ref{hamilt})
are well established in the $weak$ coupling limit.
In the limit $K \ll J_{1(2)}$, the Hamiltonian (\ref{hamilt}) describes 
a non-Haldane  spin liquid where magnon excitations are
incoherent\cite{nersesyan}.

The strong coupling regime, $K \sim J_{1(2)}$, 
has just begin to be
investigated\cite{pati,li,yamashita,frischmuth,yamashita1}. From the theoretical point of view the
situation is ackward. Indeed, as stated above, at the special point
$J_1=J_2=K/4$, the Hamiltonian (\ref{hamilt}) has an enlarged SU(4) symmetry\cite{li}
and is exactly solvable by the 
Bethe-ansatz \cite{sutherland}. The model is critical with 
$three$ gapless bosonic modes and flows in the infrared towards
the Wess-Zumino-Novikov-Witten (WZNW) model SU(4)$_1$\cite{affleck}.
It is conformaly invariant
with the central charge $c=3$.  
Therefore, there should be a $qualitative$ change in the physical behavior
of (\ref{hamilt}) when going from small to large $K$.
>From the theoretical point of view this situation is striking since 
this means that one cannot go continuously from weak to strong coupling.
This is a manifestation of the Zamolodchikov c-theorem\cite{zamolo} that states that
starting at $K=0$ with two decoupled $S=1/2$ chains with the central charge $c=2$
(two gapless bosonic modes) one cannot flow, in the Renormalization Group (RG) sense,
towards the SU(4) point which has $c=3$ (three gapless bosonic modes).
Very recently, a new approach to tackle with the strong coupling regime has
been developped by Azaria et al\cite{azaria}. The idea is to start from 
the strong coupling fixed point SU(4)$_1$ and to perturb around it.
 
This strategy has been applied to the symmetric line $J=J_1=J_2$.
It was shown that when $J<K/4$ a small deviation from the SU(4) point
is irrelevant and thus the low energy physics is  governed by the 
SU(4)$_1$ fixed point. In contrast, when $J>K/4$ the interaction is relevant and a gap
opens in the spectrum. The system dimerizes with alternating spin and orbital
singlets. In addition it was argued that the  SU(4) symmetry was restored at
long distance and that the low energy effective Hamiltonian was that of 
the SO(6) Gross-Neveu (GN) model. The low-lying $coherent$ excitations was then 
shown to be fermions that transform as an antisymmetric tensor of rank two of
SU(4). These excitations
are coherent with wavevector near $\pi/2$.

The purpose of this work is to extend the analysis to the 
asymmetric case ($J_1\!  \neq\! J_2$) and to inquire how anisotropy
modifies the low energy physics described above.  
This paper is organized as follows. In Section II we 
present the tools that are necessary to explore the vicinity of the  SU(4)$_1$
fixed point and derive the effective low-energy theory associated
with (\ref{hamilt}). A detailed renormalization group analysis is then
presented in Section III. We obtain the phase diagram
in the the plane (J$_1$, J$_2$) and discuss the asymptotic behaviors of 
the mass gap $M$. A discussion
on cross-over effects linked to the anisotropy is also presented.
In Section IV we investigate the effect of a magnetic field on the
anisotropic model. Finally in the section V we  summerize our results
and present some technical details relative to the computation of the mass gap
in the Appendix.

\section{the Low Energy Effective Field Theory}

\subsection{ The SU(4) Heisenberg chain}

Our approach is very similar to the description of the 
S=1/2 Heisenberg spin chain at low energy from the  
spin sector of the  repulsive Hubbard model at half filling\cite{affleck}.
To this end, let us introduce the SU(4) Hubbard model with $U>0$:
\bea 
{\cal H}_U &=& \sum_{ia\sigma} \left(-t \;c_{i+1 a\sigma}^{\dagger}
c_{ia\sigma} + h.c.\right) \nonumber \\
&+& \frac{U}{2} \sum_{iab\sigma\sigma^{'}} n_{ia\sigma}  n_{ib\sigma^{'}}
\left(1 - \delta_{ab} \delta_{\sigma\sigma^{'}}\right).
\label{hubbard}
\eea
Here $c_{ia\sigma}^{\dagger}$ denotes electron creation 
with channel (orbital) $a=1,2$ and spin $\sigma = \up,\down$ 
at the i-th site. The occupation number is defined by: 
$n_{ia\sigma} = c_{ia\sigma}^{\dagger} c_{ia\sigma}$. In the 
limit of large positive U, and at $quarter$ filling, it is not difficult to show that
(\ref{hubbard}) reduce to (\ref{hamilt}) with the identification:
\bea
{\vec S}_{i} &=& \frac{1}{2} \sum_a c_{ia\alpha}^{\dagger} 
{\vec \sigma}_{\alpha\beta} c_{ia\beta} \nonumber \\
{\vec T}_{i} &=& \frac{1}{2} \sum_{\alpha} c_{ia\alpha}^{\dagger} 
{\vec \tau}_{ab} c_{ib\alpha}
\label{spinrep}
\eea  
where ${\vec \sigma}$ (resp. ${\vec \tau}$)
are the Pauli matrices acting in the spin (resp. orbital) space.
The low energy physics can be described in terms of 
right movers $R_{a\sigma}$ and left movers $L_{a\sigma}$ 
fermions
which correspond to the lattice fermion $c_{ia\alpha}$
in the continuum limit:
\be 
\frac{c_{ia\sigma}}{\sqrt a_0} \simeq
R_{a\sigma}\left(x\right) \exp\left(\ri k_F x \right) + 
L_{a\sigma}\left(x\right) \exp\left(-\ri k_F x\right), \;\; x=ia_0 
\label{ccont}
\ee
where $a_0$ is the lattice spacing and 
the Fermi wavevector is defined by: $k_F = \pi/4a_0$.
 At this point we bosonize
and introduce 
four chiral bosonic fields $\Phi_{a\sigma R,L}$ 
using the Abelian bosonization of Dirac fermions\cite{tsvelik}:
\bea
R_{a\sigma} &=& \frac{\kappa_{a\sigma}}{\sqrt{2\pi a_0}}
\exp\left(\ri \sqrt{4\pi} \Phi_{a\sigma R}\right) \nonumber \\ 
L_{a\sigma} &=& \frac{\kappa_{a\sigma}}{\sqrt{2\pi a_0}}
\exp\left(-\ri \sqrt{4\pi} \Phi_{a\sigma L}\right)
\label{bosofer}
\eea
where the bosonic fields satisfy the following commutation 
relation: 
\be 
\left[\Phi_{a\sigma R}, \Phi_{b\sigma^{'} L}\right] = \frac{i}{4} 
\delta_{ab} \delta_{\sigma\sigma^{'}}
\ee
so that $\{R_{a\sigma}(x), L_{b\sigma^{'}}(y)\}=0$.
The anticommutation between fermions with 
different spin-channel indexes is
insurred by the presence of 
Klein factors (here Majorana fermions) $\kappa_{a\sigma}$
with the following anticommutation rule:
\be
\{\kappa_{a\sigma}, \kappa_{b\sigma^{'}}\} = 
2 \delta_{ab} \delta_{\sigma\sigma^{'}}.
\ee

 As in the 
the solution of the two-channel Kondo effect 
by Abelian bosonization\cite{emery} it is suitable to introduce
the physically transparent basis:
\bea
\Phi_c &=& \frac{1}{2}(\Phi_{1\up} + \Phi_{1\down} + 
\Phi_{2\up} + \Phi_{2\down})\nonumber \\ 
\Phi_s &=& \frac{1}{2}\left(\Phi_{1\up} - \Phi_{1\down} +
\Phi_{2\up} - \Phi_{2\down}\right) \nonumber \\ 
\Phi_f &=& \frac{1}{2}\left(\Phi_{1\up} + \Phi_{1\down} -
\Phi_{2\up} - \Phi_{2\down}\right) \nonumber \\ 
\Phi_{sf} &=& \frac{1}{2}\left(\Phi_{1\up} - \Phi_{1\down} -
\Phi_{2\up} + \Phi_{2\down}\right). 
\label{sfsf}
\eea
In this new basis, the total charge degree of freedom is described by 
the bosonic field $\Phi_c$ while the other ``spin-orbital'' degrees of freedom,
are faithfully  bosonized by the three bosonic fields $\Phi_s, \Phi_f, \Phi_{sf}$.
It is now straightforward to obtain the continuum limit  of 
the Hubbard Hamiltonian (\ref{hubbard}):

\bea
{\cal H}_U = {\cal H}_c + {\cal H}_{sf} 
\label{hubsepar}
\eea
where 
 
\bea
{\cal H}_c = \frac{v_F}{2} \left(\left(\partial_x \Phi_c\right)^2 
+ \left(\partial_x \Theta_c\right)^2 \right)
+ \frac{3Ua_0}{2\pi} \left(\partial_x \Phi_c\right)^2 ,
\label{chargesepar}
\eea
and
\bea
{\cal H}_{sf} = \sum_{a=s,f,sf} \left(
\frac{v_F}{2} \left(\left(\partial_x \Phi_a\right)^2
+ \left(\partial_x \Theta_a\right)^2 \right)
\right. \nonumber \\
\left.
- \frac{Ua_0}{2\pi} \left(\partial_x \Phi_a\right)^2 \right)
+  \frac{U}{\pi^2 a_0} \left(\cos\sqrt{4\pi}\Phi_s 
\cos\sqrt{4\pi}\Phi_f \right. \nonumber \\
+ 
\left.  \cos\sqrt{4\pi}\Phi_f
\cos\sqrt{4\pi}\Phi_{sf} + \cos\sqrt{4\pi}\Phi_s
\cos\sqrt{4\pi}\Phi_{sf}\right) 
 \label{spinorb}
\eea 
As in the SU(2) Heisenberg chain, spin and charge degrees of freedom separate. Notice 
however that at this order in U  there are no umklapp terms in the charge sector since we are at quarter filling
and the 4$k_F$ contribution to the effective Hamiltonian oscillates. Umklapp
terms will arise at higher order in perturbation theory and will be
responsible of a Mott transition at a finite value of $U =U_c$\cite{assaraf}.
Assuming that $U >> U_c$, we focus now on the spin-orbital sector.

The interaction term in ${\cal H}_{sf}$ has scaling dimension $2$ and is
therefore marginal. This term is nothing but the SU(4) current-current
interaction. The $15$  SU(4)$_1$ currents ${\cal J}_R^a$ and ${\cal J}_L^a$
can be expressed in terms of the three bosonic fields 
$\Phi_s$,$ \Phi_f$ and $\Phi_{sf}$ and the spin-orbital part of the
Hamiltonian (\ref{hubsepar}) takes the  form:
\be
{\cal H}_{sf} = \frac{ 2\pi v_s}{5} \sum_{a=1,15} \left( {\cal J}_R^a {\cal J}_R^a +
 {\cal J}_L^a {\cal J}_L^a \right) + 2 g_s \; \sum_{a=1,15} {\cal J}_R^a {\cal J}_L^a
\label{sugarawa}
\ee
where 
\bea
g_s &=& - Ua_0 \nonumber \\
v_s &=& v_F - Ua_0/2\pi.
\eea
The first term in Eq. (\ref{sugarawa}) is just the Sugawara form of the 
SU(4)$_1$ WZNW model while the second term is the marginal current-current interaction.
When $U>0$, $g_s<0$, the current-current interaction ${\cal J}_R^a {\cal J}_L^a$
is thus irrelevant, as a consequence the spin orbital sector is described by the
 SU(4)$_1$ WZNW model\cite{affleck}.

\subsection{Majorana representation and the SO(6) Gross-Neveu model}
 
As first emphasized by Shelton et al\cite{shelton} in their study of
the two-leg spin $1/2$ ladders it is very
convenient to formulate spin ladders problem in terms of real (Majorana) fermions.
This can be done by refermionising
the three bosonic fields $\Phi_s$,$ \Phi_f$ and $\Phi_{sf}$.
Let us introduce the six Majorana
fermions $\xi^a, a=1..6$ as follows: 
\bea
\left(\xi^1 +\ri \xi^2\right)_{R(L)} &=& \frac{\eta_1}{\sqrt{\pi a_0}}
\exp\left(\pm \ri \sqrt{4\pi} \Phi_{sR(L)}\right) \nonumber \\
\left(\xi^3 + \ri \xi^4\right)_{R(L)} &=& \frac{\eta_2}{\sqrt{\pi a_0}}
\exp\left(\pm \ri \sqrt{4\pi} \Phi_{fR(L)}\right) \nonumber \\
\left(\xi^5 + \ri \xi^6\right)_{R(L)} &=& \frac{\eta_3}{\sqrt{\pi a_0}}
\exp\left(\pm\ri \sqrt{4\pi} \Phi_{sfR(L)}\right) \nonumber \\
\partial_x \Phi_s &=& \ri \sqrt{\pi} ( \xi^1_R \xi^2_R + \xi^1_L \xi^2_L) \nonumber \\
\partial_x \Phi_f &=& \ri \sqrt{\pi} ( \xi^3_R \xi^4_R + \xi^3_L \xi^4_L) \nonumber \\
\partial_x \Phi_{sf} &=& \ri \sqrt{\pi} ( \xi^5_R \xi^6_R + \xi^5_L \xi^6_L)
\label{majorep1}
\eea
where $\eta_i$ are  Klein factors.
With all these relations at hand, one can rewrite the Hamiltonian 
(\ref{spinorb}) in terms of six Majorana fermions: 
\be
{\cal H}_{sf} = -\frac{iv_s}{2} \sum_{a=1}^{6} \left(\xi^{a}_R \partial_x \xi^{a}_R
- \xi^{a}_L \partial_x \xi^{a}_L\right) + g_s \sum_{i<j} \kappa_i \kappa_j
\label{grossneveuappendix}
\ee
where we have introduced the energy density of the different
Ising models: $\ri\epsilon_a=\kappa_a= \xi^{a}_{R} \xi^{a}_L$. The Hamiltonian
(\ref{grossneveuappendix}) is nothing but that of the SO(6) GN
model with a marginally $irrelevant$ interaction when $U>0$. Thus in the far
infrared the six Majorana fermions decouple and remain massless. This is the
SO(6)$_1$ fixed point. 

The above result assumes that $U$ is small and the question that naturally
arises is whether it can be extended to large values of $U$ where
(\ref{hubbard}) reduces to (\ref{hamilt}). For exactly the same reasons
as for the SU(2) Heisenberg chain the answer is positive. We know from the
exact solution that the SU(4) model is critical with three massless
bosonic modes or equivalently six massless Majorana fermions. We
know from Conformal Field Theory that the fixed point Hamiltonian can
only be the SU(4)$_1$ $\sim$ SO(6)$_1$ WZNW model. The marginal interaction
$\sum_{i<j} \kappa_i \kappa_j$ is the only one that respects both the SO(6)
symmetry as well as translation invariance. Therefore in  the vicinity of the
fixed point the SU(4) Heisenberg model will be given by:
\be
{\cal H} = -\frac{iu_s}{2} \sum_{a=1}^{6} \left(\xi^{a}_R \partial_x \xi^{a}_R
- \xi^{a}_L \partial_x \xi^{a}_L\right) + 2G_3 \sum_{i<j} \kappa_i \kappa_j
\label{grossneveu}
\ee
where the spin velocity $u_s$ and the coupling $G_3 < 0$ are unknown
and non universal parameter that could be extracted from the exact solution.
The only thing that happens when going from small $U$ to large $U$
is a renormalization of the $g_s$ and $v_s$.

The Majorana description used here is extremely usefull to understand the symmetry properties of
our model. Indeed for example, one can define the spin and orbital triplets:
\bea
 {\vec \xi}_{sR(L)} = \left( \xi^2, \xi^1, \xi^6\right)_{R(L)}  \nonumber \\
 {\vec \xi}_{tR(L)} = \left( \xi^4, \xi^3, \xi^5\right)_{R(L)}.
\label{stmaj}
\eea
These quantities transform like a vector under spin SO(3)$_s$ and
orbital SO(3)$_t$ rotations. These correspond to the SU(2)$_s$
and SU(2)$_t$ transformations acting on the operators 
${\vec S}$ and ${\vec T}$ respectively.

To get a complete description of the  SO(6)$_1$ fixed point one needs 
the continuum expressions for the effective spin and orbital densities
in terms of the Majorana fermions:
\bea 
{\vec S}& = &{\vec J}_{sR} + {\vec J}_{sL}  
+ \exp\left(\ri \pi x/2a_0\right) {\vec {\cal N}_s} + \rm{H.c.} + 
\left(-1\right)^{x/a_0} {\vec n}_s(x)
 \nonumber \\
{\vec T} &=& {\vec J}_{tR} + {\vec J}_{tL}
+ \exp\left(\ri \pi x/2a_0\right) {\vec {\cal N}_t} + \rm{H.c.} + \left(-1\right)^{x/a_0} {\vec n}_t(x) \nonumber \\
\label{spindensities2}
\eea
where ${\vec J}_{s,t}$ is the uniform, $k=0$, part and ${\vec {\cal N}_{s,t}}$ and
${\vec n}_{s,t}$ are the $2k_F=\pi/2a_0$ and $4k_F=\pi/a_0$ contributions.
Notice that in contrast with the SU(2) Heisenberg chain, the spin
density has three oscillating components. The  reason for this comes from Conformal Field Theory.
Indeed the different oscillating components of the spin density are $primary$ fields
of the SU(4)$_1$ WZNW model. There are three of them with scaling 
dimensions (3/4,1,3/4)\cite{difrancesco}. They all belong to the 
representations (building blocks)  of SU(4) with Young tableau consisting of $a$ (a=1,2,3)
boxes and one column. In particular the staggered components
at 4k$_F$ = $\pi$, ${\vec n}_{s,t}$, are components of a antisymmetric tensor
of rank $2$.

The uniform components express in terms of the Majorana fermions as follows:
\bea
{\vec J}_{sR(L)} = 
- \frac{\ri}{2} \; {\vec \xi}_{sR(L)} \land {\vec \xi}_{sR(L)} \nonumber \\
{\vec J}_{tR(L)} =
 - \frac{\ri}{2} \; {\vec \xi}_{tR(L)} \land {\vec \xi}_{tR(L)}
\label{curr}
\eea
Notice that in contrast with the SU(2) Heisenberg chain, the uniform part of the
spin densities are SU(2)$_2$ currents. The expressions for the 
$4k_F=\pi/a_0$ densities are given by
 \bea
{\vec n}_s = \ri B \; {\vec \xi}_{sR} \land {\vec \xi}_{sL} \nonumber \\
{\vec n}_t = \ri B \; {\vec \xi}_{tR} \land {\vec \xi}_{tL}
\label{4kFdensities}
\eea
where $B$
is a non universal constant. Their scaling dimension at the SO(6)$_1$
fixed point is $\Delta_{\pi} = 1$.
Both densities (\ref{curr}) and  (\ref{4kFdensities}) are rather simple
when expressed in terms of the Majorana fermions. This is not the case with
the $2k_F=\pi/2a_0 $ densities  ${\vec {\cal N}_{s,t}}$ that are non local
in the Majorana fermions $\xi^a$.  Indeed they involve order and disorder
operators $\sigma_a$ and $\mu_a$ of the six Ising models that are associated 
with the six Majorana fermions.
The expressions of both ${\vec {\cal N}_{s,t}}$ are  lengthly and we shall
give here only the $z$ component that will be sufficient for our purpose:
\bea
{\cal N}_{s}^z = 
A  \; \left( \ri \mu_1 \mu_2 \sigma_3 \sigma_4 \sigma_5 \sigma_6
+  \sigma_1 \sigma_2 \mu_3 \mu_4 \mu_5 \mu_6 \right) \nonumber \\
{\cal N}_{t}^z = 
A  \; \left( \ri \sigma_1 \sigma_2 \mu_3 \mu_4 \sigma_5 \sigma_6
+  \mu_1 \mu_2 \sigma_3 \sigma_4 \mu_5 \mu_6 \right)
\label{2kFdensities}
\eea
where $A$ is also a non universal constant. Since at the free fermion point,
the order and disorder operators have scaling dimension $1/8$, the
$2k_F$ densities ${\vec {\cal N}_{s,t}}$ have the scaling dimension
$\Delta_{\pi/2} = 3/4$.

This  completes the continuum description at the SU(4) point. 
The theory is critical and flows to the SO(6)$_1$ fixed
point. There is  a marginally irrelevant correction which magnitude 
is nonuniversal and  is given by the unknown coupling $G_3<0$. This is 
in complete agreement with the non-Abelian bosonization approach of 
Affleck\cite{affleck}. With   the continuum expressions of the spin and 
orbital operators at the SU(4) point in terms of the six Majorana fermions and
 the
associated Ising models we shall be  able us to investigate the properties
of (\ref{hamilt}) close to the SU(4) symmetric model. But before moving to 
this point, let us stress that the effective theory depends on three  unknown 
parameters $A$ and $B$ and $G_3$ which can be in principle extracted from 
numerical studies. In the following, we shall assume that the above 
description still holds for small deviations of the SU(4)
point. The only modification being a renormalization of the  non universal
constants $A$, $B$ and $G_3$.

\subsection{ The SU(2)$\otimes$SU(2) symmetry breaking perturbation}

We shall now derive the effective low energy theory associated with
(\ref{hamilt}) for arbitrary values of $J_{1}$ and $J_{2}$ close
to the SU(4) invariant point given by $J_{1}=J_{2}= K/4$. To this end, 
let us parametrize the couplings as follows:
\bea
J_{1} = \frac{K}{4} + G_1 \nonumber \\
J_{2} = \frac{K}{4} + G_2
\label{G1G2}
\eea
where both $G_1$ and $G_2$ are much smaller than $K$. The Hamiltonian
(\ref{hamilt}) can then be written as:

\be
{\cal H} = {\cal H}_{SU(4)} + G_1 \sum_{i} {\vec S}_{i} \cdot {\vec S}_{i+1}.
+ G_2 \sum_{i} {\vec T}_{i} \cdot {\vec T}_{i+1}.
\label{hamiltoniang}
\ee
Using the low energy description of the spin-orbital operators (\ref{curr}),
(\ref{4kFdensities}) and (\ref{2kFdensities}), one can expand
(\ref{hamiltoniang}) around the SO(6)$_1$ fixed point:

\bea
{\cal H} = &-&\frac{iu_s}{2} \left( {\vec \xi}_{sR} \cdot \partial_x {\vec \xi}_{sR} -
{\vec \xi}_{sL} \cdot \partial_x {\vec \xi}_{sL} \right) 
+ g_1 \;\left( \kappa_1 + \kappa_2 + \kappa_6 \right)^2 \nonumber \\ 
&-&\frac{iu_t}{2} \left( {\vec \xi}_{tR} \cdot \partial_x {\vec \xi}_{tR} -
{\vec \xi}_{tL} \cdot \partial_x {\vec \xi}_{tL} \right) 
+g_2 \; \left( \kappa_3 + \kappa_4 + \kappa_5\right)^2 \nonumber \\
&+& 2G_3 \; \left( \kappa_1 + \kappa_2 + \kappa_6 \right)
\left( \kappa_3 + \kappa_4 + \kappa_5 \right)
\label{hpert}
\eea
where 
\bea
g_1 &=& G_1 + G_3 \nonumber \\
g_2 &=& G_2 + G_3
\label{couplings}
\eea
and the two renormalized velocities $u_s$ and $u_t$ are given by:
\bea
u_s &=& v_s + 2 G_1 /\pi \nonumber \\
u_t &=& v_s + 2 G_2 /\pi.
\label{velocities}
\eea
In the above equations and in the remaining of this paper, we  include
the effect of  the 4k$_F$ components of the spin and orbital densities in a redefinition 
of the couplings: $G_{(1,2)} \to (1 + B^2)G_{(1,2)}$.
The Hamiltonian
(\ref{hpert}) describes two marginally coupled SO(3) Gross-Neveu models:
one in the spin channel described by the three Majorana ${\vec \xi}_{s}$
and one in the orbital channel described by the three Majorana ${\vec \xi}_{t}$.
The situation at hand is to be contrasted another time with the one
encountered in the study of spin ladders. In the latter models
deviation from criticality leads, in general, to relevant perturbation
and a gap always opens independently of the sign of the couplings. There are
however notable exceptions where frustration  plays its trick. This is the
case of some three-leg frustrated ladders where for some particular values
of the couplings only marginal interaction remains in the effective action.
As a result a non trivial critical state, the so-called "chirally stabilized"
liquid, shows off\cite{azaria2}. In the model the situation is even more striking since
we find only marginal interactions in a finite region of the couplings.
This is mainly due to the fact that the frustation is maximal in the strong
coupling region. A direct consequence of the marginality of all
the interaction is that we expect  the phase diagram to result
from a delicate balance between the different terms entering in (\ref{hpert}).
This is why it  is  now worth discussing the effect of the three different terms
in (\ref{hpert}). 

Consider first the case where $G_3=0$. 
Then we are left with the two decoupled SO(3)$_s$ and SO(3)$_t$ GN models which 
are exactly solvable. At issue is the sign of $G_1$ and $G_2$. 
When $G_1 > 0$ and $G_2 > 0$ a gap opens in both spin and orbital channels. The spectrum
consists of  kinks and antikinks (there are no fermions)\cite{zamolo2}.
When $G_1 < 0$ and $G_2 < 0$, the interaction is irrelevant and the model flows towards
the isotropic SO(6)$_1$ $\sim$ SU(4)$_1$ fixed point with the central charge $c=3$.
When $G_1G_2 < 0$  one of the  SO(3)
GN model will become massive while the other one will flow towards the 
SO(3)$_1$ fixed point. The Hamiltonian will remain critical with $three$
massless Majorana fermions  leaving the whole system with the central charge $c=3/2$.

Now the physically relevant question is 
whether or not  this scheme survives a small negative $G_3$. Indeed, 
it would be not correct to neglect the last term in (\ref{hpert}). First of all,
from the point of view of the lattice Hamiltonian (\ref{hamiltoniang}), the Hamiltonian
(\ref{hpert}) has to be thought as the effective Hamiltonian obtained by integrating  out
high energy modes up to a scale where one sits close enough to the SO(6)$_1$ fixed point
where the continuum limit can be taken. Thus, generically $G_3$ is not zero. The second reason is that  all other interactions 
are $marginal$. In such a case, it is well known that
 operators that are naively irrelevant may  become dangerous and strongly modify 
the physics in the infrared. 
Therefore, eventhough $G_3 <0$, one has to keep it and analyze with care
(\ref{hpert}) with all couplings different from zero. As we shall see,
the strong tendency to the SO(3)$_1$ criticality in the regions $G_1G_2 < 0$ 
will be spoiled in most of the parameter space. There will be though still
a finite region where the Hamiltonian will be critical but with an 
approximate SO(6) symmetry. 

\section{Renormalization Group Analysis}

\subsection{ The renormalization group flow}

The RG equations for the couplings entering in (\ref{hpert})
are given at leading  order by:
\bea
\dot{g_1} &=& \frac{1}{\pi u_s} g_1^2 + \frac{3 }{\pi u_t} G_3^2 \nonumber \\
\dot{g_2} &=& \frac{1}{\pi u_t} g_2^2 + \frac{3 }{\pi u_s} G_3^2 \nonumber \\
\dot{G_3} &=& \frac{2 G_3  }{\pi} ( \frac{g_1}{u_s} + \frac{g_2}{u_t})
\label{rg2}
\eea
In the above equation, $\dot{G}$ means $\partial G/\partial t$ where
$t=\log(\lambda)$ is the RG scale. It is more convenient to express the set of Eqs.
(\ref{rg2}) in terms of the couplings  that enter in the lattice Hamiltonian
(\ref{hamiltoniang}):
\bea
\dot{G_1} & =&  G_1^2 - 2G_2 G_3 \nonumber \\
\dot{G_2} & =&  G_2^2 - 2G_1 G_3  \nonumber\\
\dot{G_3} & =&  2G_3 ( G_1+G_2+ 2G_3 ),
\label{rgG}
\eea
where, for simplicity we have made the following redefinition:
\bea
G_1 &\to& 1/\pi\sqrt{u_s u_t} \left( \alpha G_1 - (1 - \alpha)
G_3\right) \nonumber \\
G_2 &\to& 1/\pi\sqrt{u_s u_t} \left(
 1/\alpha  G_2 - (1 - 1/\alpha) G_3\right) \nonumber \\
G_3 &\to& 1/\pi\sqrt{u_s u_t} G_3 
\label{redef}
\eea
with $\alpha= \sqrt{u_t/u_s}$. At the leading order, the velocities
$u_s$ and $u_t$ also renormalize:
\bea
\dot{u_s} & =&  - 6 u_s  G_3^2 \
{\cal I}( \frac{u_t}{u_s})  \nonumber \\
\dot{u_t} & =&     - 6   u_t  G_3^2 \
{\cal I}( \frac{u_s}{u_t})
\label{rgu}
\eea
where 
\be
{\cal I}(x) = \frac{1- x}{1+x}.
\ee
The  RG equations (\ref{rgG}) and (\ref{rgu}) can be exactly solved
and reduce to a single differential equation.
To show this let us  introduce the following variables:
\bea
 \mu &=& G_1 G_2 + G_1G_3 + G_2G_3 \nonumber \\
 d &=& G_1-G_2  \nonumber \\
 s &=& G_1+G_2+2 G_3
\label{mds}
\eea
 Eqs. (\ref{rgG}) and (\ref{rgu}) greatly simplify to
\bea
 \dot{\mu} &=& \mu  s  \nonumber \\
 \dot{d} &=& d  s  \nonumber \\
 \dot{G_3} &=& 2 G_3 s
\label{rgmsd}
\eea
and 
\bea
\dot{u_s} -  \dot{u_t} &=& - 6  G_3^2 \ (u_s - u_t) \nonumber \\
u_s u_t &=& \rm{constant}.
\label{rgdelu}
\eea
While the meaning of both  $s$ and $d$ is clear, the physical interpretation
of the variable $\mu$ is not straightforward. As we shall see in the
following, $\mu$  measures the departure from criticality. 

The solution of both (\ref{rgmsd}) and (\ref{rgdelu})  are then easily
obtained:
\bea
\mu(t) &=& \mu(0) X(t)  \nonumber \\
d(t) &=& d(0) X(t)  \nonumber \\
G_3(t)  &=& G_3(0)  X^2(t) \nonumber \\
(u_s - u_t)(t)  &=& (u_s - u_t)(0) \exp{\left( -6 \  G_3^2(0) \int_0^t
X^4(\tau) \rm{d}\tau \right)}
\nonumber \\
(u_s u_t)(t) &=& (u_s u_t)(0)
\label{sol}
\eea
where 
\be
X(t) = {\displaystyle \exp{  \int_0^t s(\tau) \rm{d}\tau }}
\label{X}
\ee
is the solution of the differential equation:
\be
\left(\frac{\dot{X}}{X}\right)^2 =  P(X).
\label{diffX}
\ee
In the latter equation $P(X)$ is a fourth order polynomial that depends on the initial
conditions of the flow:
\be
P(X) = 4 G_3^2(0) X ^4 + d^2(0) X^2 + 4 \mu(0) X.
\label{P}
\ee
Once the solution $X(t)$ of (\ref{diffX}) is known, the behavior of the RG
flow is completely determined. In particular the time evolution of
couplings $G_1(t)$ and
$G_2(t)$ is given by:\be
G_{(1,2)}(t) = \frac{1}{2} \left(  \epsilon(s) \sqrt{ P(X(t))} - 2 G_3(0) X^2(t) \pm d(0) X(t)
\right)
\label{g12t}
\ee
where $\epsilon(s) = \rm{sign}(s)$.

As in the XXZ model we distinguish $three$ different phases A, B and C
that are separated by the two surfaces defined by $\mu =0$.
\begin{figure}
	\begin{center}
	\mbox{\psfig{figure=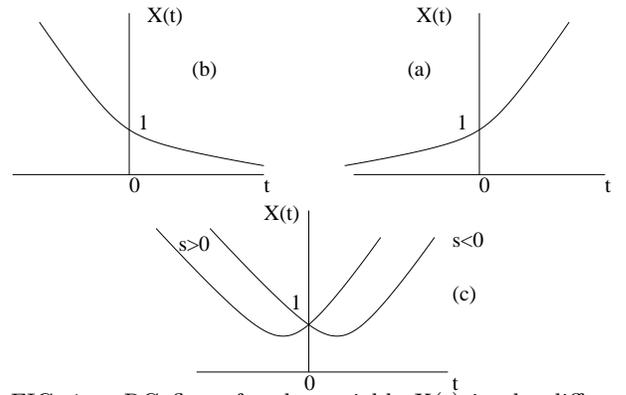,width=8cm}}
	\caption[99]{
	RG flows for the variable X(t) in the different phases A, B
and C. In the phase C, X(t) reaches a minimum value X$^*$ at a scale t$^*$.
When $s(0) < 0$, the minimum is reached for  a physical time
$t^*>0$.}
	\end{center}
\end{figure}

-{ \it The A Phase}. There one has $\mu(0) > 0$ and $s(0) >0$ and 
$X(t) \to \infty$  as $t \to \infty$ (see Fig. 1(a)). All couplings are
relevant and a gap opens in the spectrum. In addition, the velocities
$u_s(t)$ and $u_t(t)$ goes to a fixed value $u_s^* = u_t^*$. Looking in more
details in the large $t$ behavior of the couplings, one finds that 
$G_1(t) \sim G_2(t) \sim - 2 G_3(t)$ so that the effective interaction
in the far infrared takes the suggestive form: 
$  \sim  (  (\kappa_1 +  \kappa_2 + \kappa_6) - (\kappa_3 + \kappa_4 + \kappa_5)  )^2$
which after the transformation ${\vec \xi}_{sR} \to {\vec \xi}_{sR}$,
${\vec \xi}_{sL} \to - {\vec \xi}_{sL}$ acquires a manifestely
SO(6) invariant form $  \sim  ( \sum_1^6 \kappa_a )^2$. 
One may therefore be tempted to conclude  that the SO(6) symmetry is restored 
in the far infrared and that the low energy spectrum of the Hamiltonian 
would be approximatively that of the  SO(6) GN model. 
However, as pointed out by  Azaria, Lecheminant and Tsvelik\cite{azaria1},
one cannot conclude on the basis of perturbation theory alone.
In fact looking at the flow in the infrared is not sufficient and can
be missleading. The point is that in the A phase, since 
$X(t)  \to 0$  as $t \to -\infty$ all couplings goes to zero
in the  ultraviolet and the theory is asymptotically free. From
the field theoretical point of view this means that a well defined
renormalized theory with $three$ different 
renormalized couplings $G_{1R}, G_{2R}$ and $G_{3R}$
exists\cite{zinn}.  Therefore, it is most likely
that the SO(6) symmetry is $not$ restored in the A phase.
The situation here is very similar to the XXZ model where
in the A phase (the Ising region), though there is an apparent SU(2) symmetry 
restoration in the far infrared, the  $exact$ solution tells us that this is not the case\cite{jap}.

-{ \it The B Phase}. There one has $\mu(0) > 0$ and $s(0) <0$ and the flow
in the $X$ variable is reversed (Fig. 1(b)). All couplings goes to zero in the
infrared and the interaction is irrelevant.
The six Majorana fermions are massless and the  model is critical with
the central charge $c= 6 \times 1/2 = 3$. However, for generic values of the
initial conditions ($d(0) \neq 0$), the fixed point Hamiltonian does not have the SO(6)
symmetry. Indeed, the velocities in both spin and orbital sector have
different fixed point values $ u_s^* \neq  u_t^*$ so that the symmetry at the
fixed point is SO(3)$_s \otimes$ SO(3)$_t$. It is only on the symmetric line
($G_1 = G_2$) that the fixed point symmetry is SO(6).

-{ \it The C Phase}. There one has $\mu(0) < 0$ and $s$ can have both signs.
As seen in Fig. 1(c), $X(t) \to \infty$ when both
$t \to \pm \infty$. The theory is not asymptotically free neither
in the infrared $nor$ in the ultraviolet. Though one certainly expects that 
a gap opens in the spectrum, perturbation theory is pathological\cite{zinn}.
Indeed, from the field theoretical point of view,
the lack of asymptotic freedom in the ultraviolet implies that a well defined
renormalized theory with $three$ different couplings 
$G_{1R}, G_{2R}$ and $G_{3R}$ does not exist. The low energy physics
in such a situation is a highly non trivial and essentially non perturbative 
problem and one is left to make a sensible conjecture. 

As in the A phase discussed above, in the far
infrared $G_1(t) \sim G_2(t) \sim -2 G_3(t)$  as well as  $u_s^* = u_t^*$ so
one may wonder again whether the SO(6) symmetry is approximatively restored or
not. We stress that the situation at hand here is very different to what
happens in the A phase since the theory is $not$ asymptotically free in the
ultraviolet. This may have important consequences for the low energy physics.
Indeed it is well known that  the divergency of some
couplings at high energy is reminiscent of the fact that some $part$ of the
interaction is $irrelevant$ at low energy. Of course,
perturbation theory alone cannot tell us $which$ part of the interaction
is irrelevant and at present all what  we have at hand to make a 
reasonable hypothesis is our perturbative results. The simplest scenario
is that the SO(6) symmetry is approximatively restored provided the initial
conditions are not too anisotropic (see however subsection C).
As a support of this conjecture let us mention that this is what happens in the
C phase  of the XXZ  model where the Bethe ansatz solution
tells us that the SU(2) symmetry is restored up to exponentially small
corrections\cite{jap} . We are of course aware that our hypothesis is highly 
questionable but it is the simplest one and could in principle be tested
either numerically or experimentally. Indeed, 
the immediate consequence of  the possible SO(6) restoration
is that the effective low energy effective theory of the Hamiltonian 
(\ref{hamilt}) in the C phase is approximatively that of the SO(6) GN model:
\bea
{\cal H} =-\frac{iu^*}{2} \sum_{a=1}^{6} \left(\xi^{a}_R \partial_x \xi^{a}_R
- \xi^{a}_L \partial_x \xi^{a}_L\right) + G\sum_{i<j} \kappa_i \kappa_j
\nonumber \\ 
\label{grossneveuG}
\eea
where the $effective$ coupling $G$ is positive. The model (\ref{grossneveuG})
is integrable. Its spectrum is known and consists of the fundamental
fermion, with mass $M$, together with a kink of mass
$m=M/\sqrt2$\cite{zamolo2}. At this point the question that naturally arises 
is  how this enlarged SO(6) symmetry reflects in the spin and orbital 
correlation functions. The answer to this important question requires
the computation of the exact dynamical  correlation
functions in the SO(6) restored massive phase. This  could be 
accomplished in principle by the form factors approach as in the frustrated two leg ladder considered
in Ref. (\cite{allen}). However this task is even more for  the SO(6) case 
and goes beyond the scope of this paper.

\subsection{ Phase diagram}

Let us now sum up our results and present the phase diagram
associated with Hamiltonian (\ref{hpert}). As even at the SU(4) symmetric point the 
coupling $G_3$ is not zero,  the best way to visualize the phase diagram is
to fix the value of  $G_3$ and to look in the plane ($G_1$, $ G_2$). 
As one can see from Fig. 2, there are  two curves separating the three regions
A, B and C which are given by the equation:
\be
 \mu = G_1 G_2 +   G_1 G_3  + G_2 G_3 =0.
\label{curve}
\ee
In the region B all models are critical with approximate SO(6) symmetry.
The spectrum consists into six massless Majorana fermions with different velocities
in both spin and orbital sectors. As one crosses the critical line $\Sigma$
one enters in a fully massive phase (C phase) with an approximate SO(6) symmetry. 
For large positive values of the couplings $ G_1$ and  $ G_2$ one may
finally enter the region A. Notice that there is no  phase transition
between the A phase and the C phase since both phases are massive
but rather a smooth cross over. 
 \begin{figure}
	\begin{center}
	\mbox{\psfig{figure=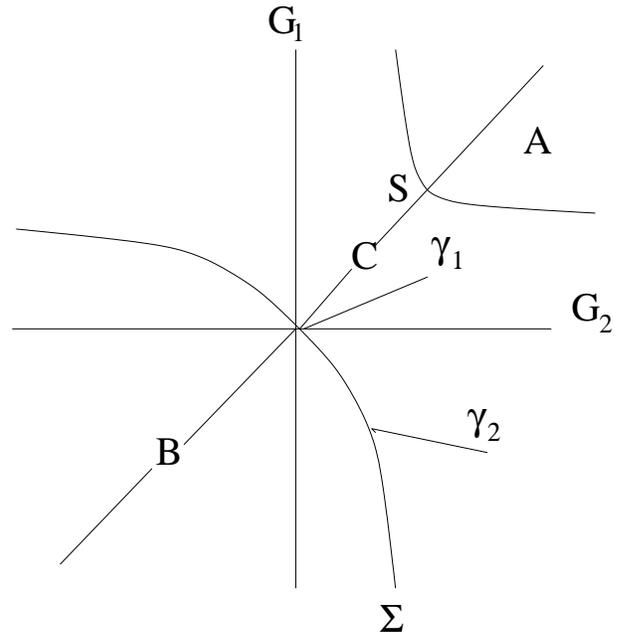,width=8cm}}
	\caption[99]{
	Phase diagram for  anisotropic couplings in the plane 
($ G_1$,  $ G_2$) at a fixed value of $G_3$. The massless phase B is
separated from the massive C phase by the critical line $\Sigma$. The special
point S on the symmetric line which is at the border between the C and the A
phases is  SO(6) symmetric with $ G_1$ =  $ G_2$ = -2 $G_3$. We plot also the
two trajectories labelled $\gamma_1$ and $\gamma_2$ for which the scaling of the mass gap 
have two different  qualitative behaviors. }
	\end{center}
\end{figure}

The most important feature of our phase diagram is that the two regions
with $G_1 G_2 < 0$ which would have been massless in the absence of $G_3$
do not survive an arbitrarily small value of 
$G_3$. Therefore the $c=3/2$ phases discussed in the previous section are not stable.
Though the whole region is essentially massive there are still room for
criticality since the B phase extends in both quadrans $G_1 G_2 < 0$. From
Eq. (\ref{curve}) the width of the critical region is of the order $|G_3|$.
Therefore the main effect of $G_3$ is to drive the system either to a fully 
massive phase (phase C) or to a fully massless phase (phase B), both 
with an approximate  SO(6) symmetry. Of course, the phase diagram as depicted
in the Fig. 2 is stricly valid in the small $G$ limit. In particular higher
order corrections may modify the shape of the critical curve $\Sigma$.
We stress however, that its behavior in the vicinity of the SU(4) point
at $G_1=G_2=0$ is given by our one loop result. In particular the fact that
$\Sigma$ crosses the symmetric line with a right angle will not change as one
includes higher order in perturbation theory. 
 However, as one goes to large deviations from
the SU(4) point our effective theory will not apply since for
large enough positive (respectively negative) $G_1$ and negative
(respectively positive) $G_2$  the orbital  (respectively spin)
degrees of freedom  order ferromagnetically\cite{pati,yamashita1}.

\subsection{ Effect of the anisotropy}

At this point one may wonder whether  the anisotropy has no effect at all in the
physical quantities. Apart from the velocity anisotropy in the B phase,
one may still expect some non trivial effect of the anisotropy in the C phase.
Indeed, after all eventhough both $c=3/2$ phases are unstable there should be
some significant signature of the presence of the SO(3)$_1$ fixed point in 
the  scaling of the physical quantities and in the finite size scaling. 
The very reason for this is that the SO(6) symmetry is restored dynamically
with help of a $marginal$ operator.

-{\it Cross-over and finite size scaling}. Let us look at the RG flow in more details. The regions of interest are those
with $G_1 G_2 <0$ in which, as discussed in the previous section, if not for
$G_3$ one of the coupling would have been irrelevant. Then  either the spin or
the orbital degrees of freedom would have remained massless. In the following
we shall concentrate on the orbital degrees of freedom. All the results that will
be given can be straightforwardly extended to the spin degrees of freedom.

Consider first initial conditions slightly above the phase B in the lower
right quadran with $G_1 G_2 <0$ (see Fig. 2). We are then in the phase C with
$G_1(0)>0$ and $ G_2(0)<0$. As seen in the Fig. 3(a), the flow can be splitted
into two "time" slices. 

\begin{figure}
	\begin{center}
	\mbox{\psfig{figure=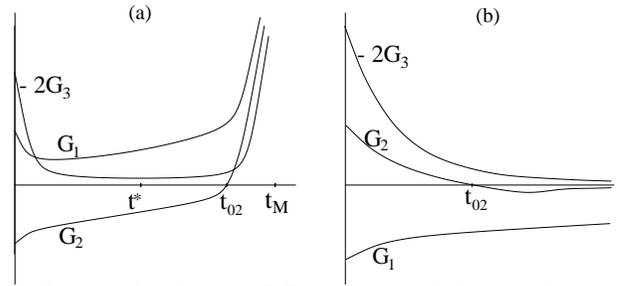,width=8cm}}
	\caption[99]{
	 Qualitative RG evolution of the coupling constants in 
regions where $G_1(0)G_2(0)<0$ and where the cross-over "time" 
defined by $G_2(t_{02})=0$ exists.}
	\end{center}
\end{figure}

At first $|G_3|$ strongly decreases and remains almost constant.
 Then,  both spin and orbital degrees of freedom weakly interact since
$|G_{(1,2)}| >> |G_3|$. In the mean time, the coupling $G_2$ increases and changes
its sign at a time $t_{02}$ where it vanishes. At "times" $t < t_{02}$ the 
orbital degrees of freedom
do not know yet they shall enter in a massive phase and the system is on the
influence of the SO(3)$_1$ fixed point. 
Finally, as $t \to t_M$ all the couplings blow 
up  and one enters in the strong coupling region. The physical interpretation
of this phenomenon is clear: it is the spin degrees of freedom that 
drive the orbital degrees of freedom  away from criticality 
and it takes roughly $\exp(t_{02})$
RG iterations for the orbital sector to escape from the bassin of attraction
of the SO(3)$_1$ fixed point. 

The other interesting region is the portion of 
the B phase delimited by the critical curve $\Sigma$ and the $G_2$-axis 
in upper left quadran of Fig. 2. There $G_1(0)<0$ and $ G_2(0)>0$.
As seen in Fig. 3(b),  it is the spin degrees of freedom 
that drive the orbital ones to criticality. The coupling $G_2$ changes its sign
and vanishes at a time $t_{02}$ and goes to zero in the limit 
$t \to \infty$.
In both cases, the cross-over "time" $t_{02}$ is given by the implicit equation:
\be
X(t_{02}) =\sqrt{\frac{\mu(0)}{d(0) G_3(0)}}.
\ee

Of course a similar discussion holds for the coupling $G_1$ and defines
another cross-over time $t_{01}$ which is defined by the implicit equation:
\be
X(t_{01}) =\sqrt{\frac{-\mu(0)}{d(0) G_3(0)}}.
\ee

We plot in Fig. 4 the iso-$t_{02}$ and iso-$t_{01}$curves in the plane ($G_1$,$G_2$).
 \begin{figure}
	\begin{center}
	\mbox{\psfig{figure=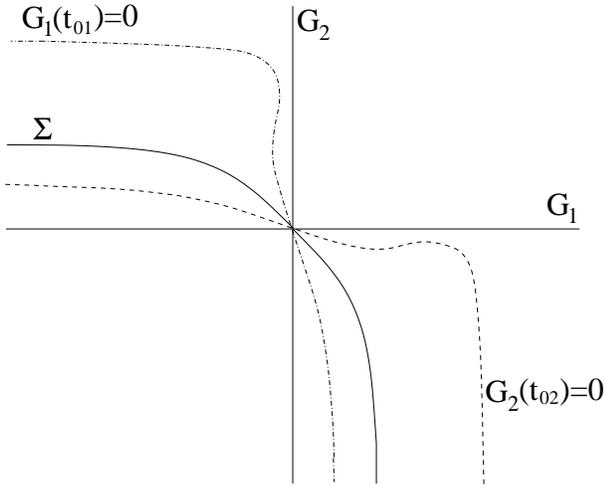,width=8cm}}
	\caption[99]{
	 Iso-$t_0$ curves in the plane ($G_1$, $G_2$); the dashed line corresponds 
to an iso-$t_{02}$ associated with the orbital degrees of freedom whereas the dashed doted
 line is an iso-$t_{01}$ associated with the spin degrees of freedom. $\Sigma$ denotes
 the critical line.}
	\end{center}
\end{figure}

We stress that this cross-over phenomenon may have important practical
consequences in numerical simulations. 
Indeed, in a finite system of size $L$, the critical region will seem
to extend
towards the iso-$t_{01}$ with $t_{01}  = \ln L$ in the region $G_1 < 0$, $G_2 >0$ and 
towards the iso-$t_{02}$ with $t_{02} = \ln L$ in the region $G_1 >0$, $G_2 <0$. The low
lying spectrum will look like if either the spin or the orbital degrees
of freedom would have been massless. It is important to notice that 
these two pseudo-critical lines bend upwards in contrast with the critical
line $\Sigma$.

-{\it The mass gap}. The anisotropy has also non trivial effect on the mass
gap of the system. To see this we have computed the mass gap
$M$ with help of the RG equations (\ref{rgG}). As well known, the gap $M$
is defined as the scale where perturbation theory breaks down. More
precisely it is given by the scale $t_M = \ln (1/Ma_0)$ at which the couplings 
blow up:
\be
X(t_M )=  \infty.
\label{gapX}
\ee
The above implicit equation can be solved but the resulting analytical
expression is too cumbersome to be quoted here. Details are given in the
Appendix  and we shall content ourselves by  its asymptotic behaviors as one
approaches the critical surface $\Sigma$.
 \begin{figure}
	\begin{center}
	\mbox{\psfig{figure=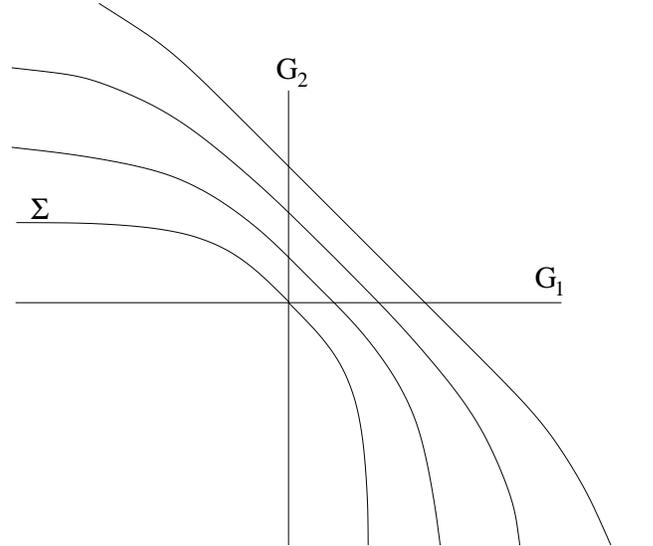,width=8cm}}
	\caption[99]{
	 Isogap curves in the plane ($G_1,G_2$) ; $\Sigma$ denotes the critical line.}
	\end{center}
\end{figure} We find two qualitatively
different behaviors depending on the way one approaches $\Sigma$.
When one approaches $\Sigma$ at the SU(4) point ($G_1 = G_2  = 0$) we find:
\be
M \sim \Lambda \exp{(- C(\gamma_1)/\Delta^{2/3})}
\label{smallgap}
\ee
where $\Delta = \sqrt{ G_1^2 + G_2^2} \to 0$ and the constant
$C(\gamma_1)$ depends on the trajectory labelled $\gamma_1$ in the Fig. 2.
On the other hand, as one approaches $\Sigma$ at any other point
the asymptotic behavior of the gap is different:
\be
M \sim \Lambda \exp{(- C(\gamma_2)/\Delta)}
\label{largegap}
\ee
where now  $\Delta$ is the Euclidean distance of the critical surface.
As above $C(\gamma_2)$ depends on the curve  labelled $\gamma_2$ in Fig. 2.
There are two other marginal cases   as one approaches $\Sigma$ 
tangentially: the exponents of $\Delta$ in both (\ref{smallgap}) and
(\ref{largegap}) are doubled. We see that the gap is generically $larger$ 
in the regions with $small$ anisotropy. To visualize this phenomenon we 
present in the Fig. 5 the isogap curves in the plane ($G_1, G_2$). 
We observe that at a given distance $\Delta$ from the critical line $\Sigma$
the gap increases as one moves towards the symmetric region
($ G_1 \sim G_2$) and is maximum on the symmetric line ($G_1 = G_2$).
Similarly, to keep the value of the gap to some constant $M$  one has to 
move away from $\Sigma$ as one leaves the symmetric region. 

-{\it Effect of anisotropy on the SO(6) symmetry restoration}. We conclude
this section by looking at the effect of the anisotropy on the SO(6) symmetry
restoration in the C phase. As discussed above, in the C phase the couplings
$G_1$ and $G_2$ tend to $-2 G_3$ in the far infrared (i.e. as $t \to
t_M$) so that the effective Hamiltonian has an apparent SO(6) symmetry.
Looking in more details at the asymptotic behavior of the couplings 
$G_1(t)$ and $G_2(t)$ we find that:
\be
G_{(1,2)}(t) \sim - 2 G_3(0) X^2(t) \pm \frac{d(0)}{2} X(t).
\label{g12as}
\ee
Therefore, as when $t \to t_M$, $X(t) \to \infty$,
there remains a subdominant infrared singularity when $d(0) \neq 0$. We
thus expect some corrections to the SO(6) behavior. To get an estimate 
of these corrections we make the reasonable hypothesis
that, when the anisotropy is small enough, the effective theory will be given by an SO(6)
GN model with a small symmetry-breaking perturbation proportional to:
\bea
{\cal H} \sim-\frac{iu^*}{2} \sum_{a=1}^{6} \left(\xi^{a}_R \partial_x \xi^{a}_R
- \xi^{a}_L \partial_x \xi^{a}_L\right) + G\sum_{i<j} \kappa_i \kappa_j
\nonumber \\ 
+ \lambda \left( \left( \kappa_3 + \kappa_4 + \kappa_5\right)^2 - 
\left( \kappa_1 + \kappa_2 + \kappa_6\right)^2 \right),
\label{grossneveuGani}
\eea
where $\lambda  << G$. Of course we do not know the value of the effective
coupling $\lambda$, but it is sufficient for our purpose that it is small
enough which will be the case if $d(0)$ is small. From 
Eq. (\ref{grossneveuGani}) one can get an estimate of the mass splitting
in both the spin and orbital sectors:
\be
\frac{M_s}{M_t} \sim 1 +  \frac{\lambda}{\ln{(1/Ma_0)}}.
\label{massplit}
\ee
Therefore, we expect  the SO(6) symmetry will
hold approximatively only  for small enough gap and anisotropy. The above
argument is of course far from being rigorous but it helps to get an idea
of the effect of the anisotropy in the symmetry restoration process.
This should serve as a warning that, once again, one  should be carefull 
with the conclusions drawn from perturbation theory.

\section{Effect of a magnetic field}

In contrast with the SU(2) Heisenberg chain there is no unique way
to apply a magnetic field in the SU(4) model. Indeed, while in the SU(2)
case there is, apart from the total spin,  only one conserved charges S$^z$
in the SU(4) model there are $three$ of them. These are the three Cartan 
generators
of the SU(4) Lie algebra. In this work we have chosen for the three commuting
generators:
\be
 H^1 = S^z, \ \ H^2= T^z \ \ \rm and \ \ H^3=2 S^z T^z.
\label{cartan}
\ee
Other choices are possible but (\ref{cartan}) is the one that is physically
relevant to our problem. We thus see that one is at liberty to apply a
magnetic field in any "direction" in the Cartan basis  (\ref{cartan}).
However, if one relies on the physical interpretation of both operators
$\vec{S}$ and $\vec{T}$ as spin and orbital operators, a  magnetic
field  only couples to $ H^1 = S^z$. The resulting Hamiltonian thus writes:
\bea 
{\cal H}_h = \sum_{i}   J_1 \; {\vec S}_{i} \cdot {\vec S}_{i+1}
+ J_2 \; {\vec T}_{i} \cdot {\vec T}_{i+1} \nonumber \\
+ K\left({\vec S}_{i} \cdot {\vec S}_{i+1}\right)
\left({\vec T}_{i} \cdot {\vec T}_{i+1}\right) +  h  \sum_{i} S^z_i.
\label{hamilth}
\eea

The magnetic field breaks the SU(4) symmetry down to 
SU(2)$\otimes$ U(1). 
For a sufficiently large value of $h > h_o$ 
the spin degrees of freedom align  parallel to the field and the remaining
degrees of freedom will decouple. More interesting is the situation
when  $h$ is small. In this case, one may expand (\ref{hamilth}) around
the SU(4)$_1$ fixed point, as in previous section, and study its effects
in perturbation. For small values of $h$, the effective low energy
Hamiltonian associated with  (\ref{hamilth}) can be easily derived using 
the expression of the spin operator $S^z$ in terms of the Majorana fermions
as given by Eqs. (\ref{curr}),(\ref{4kFdensities}) and (\ref{2kFdensities}).
Dropping all oscillating terms, the only contribution comes from
the uniform part of the spin density and the effect of the magnetic field is
just to add to the Hamiltonian (\ref{hpert}) the interaction
$ {\cal H_Z}= \ri h ( \xi_R^1 \xi_R^2 + \xi_L^1 \xi_L^2)$ and  the low-energy
Hamiltonian thus writes:
\bea
{\cal H}_h = &-&\frac{iu_s}{2} \left( {\vec \xi}_{sR} \cdot \partial_x
 {\vec \xi}_{sR} -
{\vec \xi}_{sL} \cdot \partial_x {\vec \xi}_{sL} \right) 
+ g_1 \;\left( \kappa_1 + \kappa_2 + \kappa_6 \right)^2 \nonumber \\ 
&-&\frac{iu_t}{2} \left( {\vec \xi}_{tR} \cdot \partial_x {\vec \xi}_{tR} -
{\vec \xi}_{tL} \cdot \partial_x {\vec \xi}_{tL} \right) 
+g_2 \; \left( \kappa_3 + \kappa_4 + \kappa_5\right)^2 \nonumber \\
&+& 2G_3 \; \left( \kappa_1 + \kappa_2 + \kappa_6 \right)
\left( \kappa_3 + \kappa_4 + \kappa_5 \right)  \nonumber \\
 &+& \ri h ( \xi_R^1 \xi_R^2 + \xi_L^1 \xi_L^2)
\label{hpertmag}
\eea
where we recall that
\bea
g_1 &=& G_1 + G_3 \nonumber \\
g_2 &=& G_2 + G_3.
\label{couplings}
\eea

\subsection{Renormalization group analysis}

The situation at hand is similar to that of the XXZ model in a
field\cite{giam}. There exists a magnetic length
$\lambda_h$ below which the magnetic field has no substential effect
on the physics. However as the scale $\lambda >> \lambda_h$, the field
strongly modifies the low energy behavior. Indeed, in this regime, as we shall see
below, the two Majorana fermions $\xi^1$ and $\xi^2$ completely decouple from
the four others. The best way to see this is to bosonize the "1-2" sector of the Hamiltonian
(\ref{hpertmag})  with help of Eq.(\ref{majorep1}). The resulting Hamiltonian
becomes:
\bea
{\cal H}_h &=& \frac{v_{\phi}}{2} \left( \frac{1}{K} \ \ 
\left(\partial_x \Phi_s\right)^2
+ K \ \ \left(\partial_x \Theta_s\right)^2 \right)
\nonumber \\
&-&  \frac{iv_6}{2} 
\left(  \xi^6_{R}  \partial_x  \xi^6_{R} -\xi^6_{L}  \partial_x  \xi^6_{L} \right)
\nonumber \\
&-&\frac{iv_t}{2}   
\left( \vec{\xi}_{tR} \cdot  \partial_x  \vec{\xi}_{tR} 
-\vec{\xi}_{tL}  \cdot \partial_x \vec{\xi}_{tL}   \right) \nonumber \\
&+& 2 f_2 \; \left( \kappa_3  \kappa_4 + \kappa_3 \kappa_5 +  \kappa_4 \kappa_5 \right) 
\nonumber \\
&+& 2 f_3 \;\kappa_6 \left( \kappa_3 + \kappa_4 + \kappa_5 \right) \nonumber \\
&-& \frac{2 i f_4}{\pi a_0} \cos\left( \sqrt{4 \pi} \Phi_s + 2q x \right)  
\left( \kappa_3 + \kappa_4 + \kappa_5\right) \nonumber \\
&-& \frac{2 i f_5}{\pi a_0} 
\cos\left( \sqrt{4 \pi} \Phi_s + 2q x \right) \kappa_6
\label{hmagboso}
\eea
where 
\bea
K &=& 1 - \frac{g_1}{\pi u_s}  \nonumber \\
v_{\phi} &=& u_s + \frac{g_1}{\pi }  \nonumber \\
q &=& - \frac{K h}{v_{\phi}}.
\eea
In the above Hamiltonian, we have introduced  new velocities and coupling
constants with bare values: $v_6 = u_s$, $v_t = u_t$, $f_2 = g_2$, $f_3 = f_4= G_3$ 
and $f_5 = g_1$ corresponding to the operators that are stable under 
renormalization. As readily seen,  the cosine terms   involving $\Phi_s$ will
start to oscillate with wavevector $q \sim h$. Therefore,
at sufficiently large magnetic field the two last
terms in (\ref{hmagboso}) can be dropped and the bosonic field
$\Phi_s$ completely decouples.  When a gap is present, i.e. when on sits
in the C phase at $h=0$, there exists a critical field $h_c \sim M$
below which the field has little effects on the low energy physics.
As $h$ increases and
becomes greater than $h_c$, the spin sector undergoes a
commensurate-incommensurate transition of the JNPT type\cite{jnpt}
to an incommensurate massless phase with central charge $c_s =1$. In contrast,
in the absence of a gap, i.e. when one sits in the B phase at $h=0$,
the spin sector always decouples  at sufficiently low energies.
Of course, this decoupling procedure has to be understood in the framework
of the renormalization group. To this end we have computed 
the one loop recursion relation associated
with the Hamiltonian (\ref{hmagboso}). Neglecting the velocity renormalization
we obtain:
\bea
{\dot K} &=& - (3 f^2_4 + f^2_5)J_0(2q) \nonumber \\
{\dot q} &=& q - (3 f^2_4 + f^2_5)J_1(2q) \nonumber \\
{\dot f_2 } &=&  f^2_2 + f^2_3 + 2 f^2_4 J_0(2q) \nonumber \\
{\dot f_3} &=& 2  f_2 f_3 + 2  f_4 f_5  J_0(2q) \nonumber \\
{\dot f_4} &=& (1 - K)f_4 + (2  f_2 f_4 + f_3 f_5) (1 + J_0(2q))/2  \nonumber \\
{\dot f_5} &=& (1 - K)f_5 + 3  f_3 f_4 (1 + J_0(2q))/2
\label{rgh}
\eea
where the couplings have been rescaled as $f \to  f/\pi v^*$,
 $v^*$ being a velocity scale, and $J_{(0,1)}(x)$ are the Bessel functions
of the first and second kinds. Apart from the inherent anisotropy in our
problem, the above equations resemble to that of the XXZ model in a field
\cite{giam}. We see  on Eqs. (\ref{rgh}) that when $q(t) \sim \pi$
 the  two Bessel functions  start to oscillate and the renormalization
comming from both the $f_4$ and $f_5$-terms is stopped. This defines
the magnetic length $\lambda_h$ as $q(t_h) = \pi$ ($t_h = \ln \lambda_h$)
where the spin degrees of freedom decouple from the rest of the interaction
(as well known, there is no unique way to define the magnetic length but we
have checked that other reasonable choices do not affect  qualitatively 
the physics). At this point, it is worth stressing that in the massive phase,
$t_h$ must be much smaller than $t_M$, the scale at which perturbation 
theory breaks down. This means,
of course, that $h > M$ as discussed above.

At times $t > t_h$, the couplings $f_4$ and $f_5$ do not
participate in the  RG equations of the couplings $K$, $q$, $f_2$ and $f_3$
and Eqs.(\ref{rgh}) reduce to:
\bea
{\dot K} &=& 0 \nonumber \\
{\dot q} &=& q 
\label{rgspin}
\eea
and
\bea
{\dot f_2} &=&  f_2^2 + f_3^2  \nonumber \\
{\dot f_3 } &=&  2 f_2 f_3.
\label{rg}
\eea
This means that in the regime $t > t_h$ the Hamiltonian (\ref{hmagboso})
decouples:
 \be
{\cal H}_h = {\cal H}_{s} + {\cal H}_{\perp}
\ee
where
\be
{\cal H}_{s} = \frac{v_{\phi}}{2} \left( \frac{1}{K} \ \ 
\left(\partial_x \Phi_s\right)^2
+ K \ \ \left(\partial_x \Theta_s\right)^2 \right)
 \label{hspinmag}
\ee
and
\bea
{\cal H}_{\perp} &=& - \frac{iv_6}{2} 
\left(  \xi^6_{R}  \partial_x  \xi^6_{R} -\xi^6_{L}  \partial_x  \xi^6_{L} \right)
\nonumber \\
&-&\frac{iv_t}{2}   
\left( \vec{\xi}_{tR} \cdot  \partial_x  \vec{\xi}_{tR} 
-\vec{\xi}_{tL}  \cdot \partial_x \vec{\xi}_{tL}   \right) \nonumber \\
&+& 2 f_2 \; \left( \kappa_3  \kappa_4 + \kappa_3 \kappa_5 +  \kappa_4 \kappa_5 \right) 
\nonumber \\
&+& 2 f_3 \;\kappa_6 \left( \kappa_3 + \kappa_4 + \kappa_5 \right) \nonumber \\
\label{hSO4pert}
\eea
where  $K$,$v_{\phi}$, $v_6$, $v_t$, $f_2$ and $f_3$ are the
$effective$ couplings at the magnetic length $\lambda_h$.

Let us first concentrate  on the spin sector as given by (\ref{hspinmag}).
The Hamiltonian (\ref{hspinmag}) is that of a Luttinger Liquid
with stiffness $K$. The spin sector is thus massless and contribute to a
central charge $c_s =1$. In addition, the correlation functions
involving the field $\Phi_s$ will be incommensurate 
with $h$-dependent wavevector $q$. From Eqs. (\ref{2kFdensities})
one sees that the incommensurability will reflect in both the spin and orbital
correlation functions.

We focus now on the remaining part of the interaction. The Hamiltonian
${\cal H}_{\perp}$ describes the interaction between  the orbital sector 
with the remaining spin-orbital Z$_2$ degree of freedom ($\xi^6$) of
the spin sector. The low energy physics 
in this  sector is non trivial and at issue is the 
behavior of the RG flow associated with Eq. (\ref{rg}) where 
the initial conditions have to be taken at the
magnetic length with $f_2(t_h)$ and $f_3(t_h)$ at $t_h= \ln(\lambda_h)$.
 
These equations are trivially solved. Indeed, upon introducing the new
variables:
\be
f_{\pm} = f_2 \pm f_3
\label{gpm}
\ee
Eqs.(\ref{rg}) decouple:
\be
{\dot f_{\pm}} =   f_{\pm}^2.
\ee
As in the previous section, we distinguish between three phases A, B and C
depending on the initial conditions of the flow:

- {\it The A Phase}. This is when $f_{+}(t_h) >0$ and  $f_{-}(t_h) >0$.
Both couplings are relevant and a gap opens in the spectrum. Moreover,
since the theory is asymptotically free in the ultraviolet there are two length scales
in the problem: $m_{\pm} \sim \exp{- (\pi/ f_{\pm}(t_h))}$.

- {\it The B Phase}. There $f_{+}(t_h) <0$ and  $f_{-}(t_h) <0$. 
The couplings are irrelevant and the four Majorana fermions become massless
leaving the theory with the central charge $c_{\perp}=2$. The fixed point
has  only an approximate SO(4) symmetry since there remains a
velocity anisotropy. Indeed, as in the zero field case, in general
$v_t^* \neq v_6^*$. The generic symmetry of the fixed point is thus
rather SO(3) $\otimes Z_2$.

- {\it The C Phase}. Finally is  the C phase  where 
$f_{-}(t_h) >0$ and  $f_{+}(t_h) <0$.
Then $f_{-}(t)$ is relevant and 
$f_{+}(t)$ will go to zero in the infrared. Therefore, as in the previous
section, one may conjecture that the SO(4) symmetry is approximativelly
restored. In the far infrared the effective Hamiltonian is
 that of the SO(4) Gross-Neveu model:
\bea
{\cal H}_{\perp} =-\frac{iu}{2} \sum_{a=3}^{6} \left(\xi^{a}_R \partial_x \xi^{a}_R
- \xi^{a}_L \partial_x \xi^{a}_L\right) +
 f_{-}(t_h) \sum_{i<j} \kappa_i \kappa_j
\nonumber \\ 
\label{grossneveuSO4}
\eea
which is integrable. Its spectrum consists only on kinks and antikinks 
with mass  $m \sim \exp{- (\pi/ f_{-}(t_h))}$\cite{zamolo2}.

Notice, and this will be important for the discussion that will
follow, that in both the phases B and C the effective theories
are given (up to a velocity anisotropy)  by (\ref{grossneveuSO4}) with 
the difference that  $ f_{-}(t_h)$ is $negative$ in the phase B (so that the
interaction is irrelevant) while it is $positive$ in the phase C.

\subsection{Phase diagram}

It is clear from the discussion given above that the values
of the effective couplings  at the magnetic length are crucial.
Of course, the existence of the commensurate-incommensurate
transition in the spin sector depends only on the mass gap $M$
of the zero field Hamiltonian. What is more interesting,
is what happens in the remaining orbital and spin-orbital sectors
described by the Majorana fermions $\xi^3,\xi^4,\xi^5,\xi^6$. As we shall
see  now the anisotropy will  play its tricks. Indeed, what is into question
is the sign of the coupling $f_{-}(t_h)$ $at$ the magnetic length. Returning
to the original variables, one finds that $f_{-} = G_2$. We saw in the 
preceeding section that in zero magnetic
field the time evolution  of $G_2$ was very sensitive to the presence
of the SO(3)$_1$ fixed point in the orbital sector and could change its sign
at a time $t_{02}$ depending on the the initial conditions 
(see Fig. 3(a) and Fig. 3(b)).
The presence of a field
$h$ does not affect qualitativelly this feature and provides for a
renormalization of $t_{02}$ which becomes $h$-dependent: $t_{02} \to
t_{02}(h)$. We consider now two cases. 

- First is when one sits in the B phase 
between the $G_1$ axis and the critical surface $\Sigma$, with $G_1<0$ (see Fig. 2). 
At zero magnetic field the system is critical with the central
charge $c=3$. There $f_{-}$ is positive  and decreases as $t$ increases vanishes
at some RG time $t_{02}(h)$ and then changes sign. Now if $t_h < t_{02}(h)$,
$f_{-}(t_h)$ will be still positive and a gap will open in the orbital
and spin-orbital sector according to (\ref{grossneveuSO4}). On the other hand,
if $t_h > t_{02}(h)$ then $f_{-}(t_h) < 0$ and there is no gap. This means
that there exists a critical value of the field $h_{c0}$ above which
a gap opens. The portion of
critical surface $\Sigma$ in the region $G_1 < 0$ is thus unstable.
The physical interpretation of this result is clear. In zero field, it was 
the spin degrees of freedom that drived the orbital degrees of freedom
to criticality. When the field is large enough, its effect is to decouple
a part of the spin degrees of freedom $before$  the remaining fluctuations
had a chance to enter the bassin of attraction of the SO(4)$_1$ fixed point. 
Thus, the effect of the magnetic field in this region is to reduce the 
extension of the phase B. 

- The other interesting region, is when one sits in the C phase just
above the critical surface $\Sigma$ in the lower right quadran of Fig. 2,
i.e. when $G_1 >0$ and $G_2 <0$. There $G_2$ is negative but
is driven to positive values by the spin degrees of freedom. It changes 
sign at a time $t_{02}(h)$ where it vanishes. Now if $t_h < t_{02}(h)$,
$f_{-}(t_h)$ will be  still negative while if $t_h > t_{02}(h)$, $f_{-}(t_h)$
will be positive. Therefore there exists a critical field
$h_{c0}$ below  which the orbital and spin-orbital sectors
will be still massive. Above $h_{c0}$, the gap will close.
Again, the physical reason why the gap vanishes above $h_{c0}$
is that, the spin degrees of freedom did not
have enough time to drive the orbital sector to strong coupling.
We therefore conclude that the B phase has a tendency to extend in the region
$G_1 > 0$, $G_2 < 0$ when a field is present.

To summerize, we expect two kinds of transition as one varies the
magnetic field. When the theory is massive  at $h=0$,
as one increases $h$ there will be a first transition in the spin sector
to an incommensurate phase with the central charge $c_s =1$.
The remaining degrees of freedom will be still massive
but are described by the SO(4) GN model. The coherent fermionic excitations
of the SO(6) GN model disapear from the spectrum and the only massive
excitations that remain are  the kinks of the SO(4) GN model. Consequently
all excitations will be incoherent. 
What happens as one increases $h$ further strongly depends on the anisotropy. 
If $G_2 > 0$ the magnetic field just renormalizes the mass of the
SO(4) kinks, the spectrum is still massive. The total central
charge of the model is thus $c= c_s + c_{\perp} = 1$.
However, when  $G_2 < 0$, there is a $second$ phase transition
at a field $h_{c0}$ of the Kosterlitz-Thouless (KT) type\cite{itoi1} to a 
commensurate massless phase with $c_{\perp} = 2$.  Both spin and orbital
degrees of freedom are massless and the total central charge is
$c= c_s + c_{\perp} = 3$.

Notice that there will
be three different velocities: $v^*_6 \neq v^*_{\phi} \neq v^*_t$
so that the symmetry at the fixed point is not SO(6) but rather
SO(3)$\otimes$ U(1)$\otimes Z_2$. 

When there is no gap at zero field
the spin sector is always critical with incommensurate correlation
functions. What happens for the  spin-orbital and orbital degrees
of freedom depends again strongly on the anisotropy. When
$G_2 <0$ they  remain massless and the total central charge
is thus $c=3$. However, if $G_2 >0$, there will be a KT type phase
transition at a critical field $h_{c0}$ to a $massive$ phase
with approximate SO(4) symmetry.

We stress that the mechanism that leads to the
KT type phase transition at the magnetic field $h_{c0}$
is highly non trivial since the magnetic field does $not$
couple directly  to both the orbital and the spin-orbital degrees of freedom.

\section{Conclusions}

In the present work we have studied the effect of symmetry breaking
perturbations in the one dimensional SU(4) spin-orbital model. Using the 
low energy effective field theory developped in Ref.\cite{azaria},
we have investigated the phase diagram of the SU(2)$\otimes$SU(2)
model where the exchange in both the spin ($J_1$) and the orbital ($J_2$)
sectors are different. We found that the different 
phases of the symmetric $J_1 =J_2$ line extend to the case $J_1 \neq J_2$.
In particular the massless phase, governed by the SO(6)$_1$
fixed point, extends to a finite region in the plane $(J_1, J_2)$
around the SU(4) point  $(J_1=K/4, J_2=K/4)$.
 Similarly, in the vicinity
of the critical surface, the massive phase has also an 
approximate SO(6) symmetry provided the anisotropy is not too large.
In this phase, as in the isotropic case, the system spontaneously breaks 
translational invariance and dimerizes with alternate spin and orbital 
singlets\cite{azaria}. Both spin and orbital excitation are coherent at wave vector 
$k=\pi/2$. All these results remain valid in the vicinity of the SU(4)
point. The question that naturally arises is what happens when $K$ decreases.
Indeed, in the limit $K \ll J_{(1,2)}$ one enters in the weak coupling limit
where magnon excitations are incoherent at wavevector $k=
\pi$\cite{nersesyan}. In the simplest
scenario, as discussed in  Ref.\cite{azaria}, one expects that the coherent
peak at $k=\pi/2$ in the dynamical susceptibility will disapear at a
critical value of $K=K_D$. Such a special point where an oscillating component
of the correlation function disapear is a disorder point\cite{stephenson}
and therefore, we do not expect a phase transition at $K_D$
but rather a smooth cross over.

Though these results could have been anticipated on the basis of the previous
study of the symmetric case\cite{azaria}, since the interactions
are marginal,  the anisotropy reveals itself in a non trivial scaling of the 
physical quantities. Indeed, we have shown that  the anisotropy plays its 
tricks in two particular regions of the phase diagram with 
$G_1 G_2 < 0$, where $G_{(1,2)} = J_{(1,2)} - K/4$ measures the departure
from the SU(4) point. In these regions,  both spin and orbital degrees
of freedom compete. For instance, when $G_1 >0$ and  $G_2 <0$, 
the spin sector tends to open a gap while the orbital one wants
to remain massless. Since both sectors interact marginally, at issue
is a delicate balance between the strength of the interactions:
it is one kind of degrees of freedom that drives
the other to its favorite behavior. In particular, 
this is the reason why the massless phase extends in the region with
either $G_1 >0$ or $G_2 >0$. In any cases, the whole system ultimately 
becomes either fully  massless or  fully massive.  The crucial point
is that since the interactions are only marginal, 
it may take a very long time, in the renormalization group sense, to reach
the asymptotic low energy regime. This has important consequences.

First of all is the non trivial behavior of the mass gap of the system.
We found that the gap $M$ is genericaly smaller in the regions
with large anisotropy, i.e. in the two quadrans $G_1 G_2 < 0$ above
the critical curve $\Sigma$. This is due
precisely to the strong tendency to massless behavior in these regions.  
As a consequence the gap $M$ has two qualitatively different scaling
behaviors as one approaches $\Sigma$ either from the symmetric region
or the asymmetric one (the trajectories labelled $\gamma_1$ and
$\gamma_2$ in Fig. 2): $M \sim \exp(-C_1/\Delta^{2/3})$ and 
$M \sim \exp(-C_2/\Delta)$.

Second is the finite size scaling. Since the gap opens exponentially
it is very difficult to localize accurately the critical
line $\Sigma$ in a finite system. In the current model the situation
is even more ackward  in the regions $G_1 G_2 < 0$.
Indeed, in a finite system of size $L$, 
the critical region will seem to extend and the pseudo-critical lines
will be given by the two iso-$t_{01}$ and iso-$t_{02}$ curves,
with $t_{0(1,2)}  = \ln L$, that have the opposite curvature than
the true critical line $\Sigma$ (see Fig. 4.). In this pseudo-critical region,
either the spin or the orbital degrees of freedom will look massless.
The phase diagram in zero magnetic field as obtained by 
very recent DMRG  calculations\cite{yamashita1,itoi}  is in
qualitative agreement with our RG analysis. However,
the critical line obtained in these numerical computations has the opposite 
curvature that the one loop result $\Sigma$. Our interpretation of this
fact is that   what has been observed are the two iso-$t_{0(1,2)}$ curves.
This reflects once again the non trivial finite size scaling induced by the
anisotropy.

Finally, is the effect of a magnetic field. 
The magnetic field affects the spin degrees of freedom in the usual
way. In the massless phase it leads to incommensuration in the spin
sector while when a gap is present, a commensurate-incommensurate transition
can occur at a critical field. However, what happens to the remaining
degrees of freedom strongly depends on the anisotropy.
In the region, where both degrees of freedom do not compete, i.e. 
when $G_1 G_2 > 0$, the remaining orbital and spin-orbital sector
remains either massless or massive with an approximate SO(4) symmetry.
On the other hand, the most stricking effect occurs when the spin and 
orbital fluctuations
compete i.e.  for $G_1 G_2 < 0$. In this region the field 
reinforces the effect of the orbital degrees of freedom and 
can induce a second  phase transition, of the KT type,
for a sufficiently large field, from massive to massless approximate SO(4)
behavior. The origin of this non trivial effect of the magnetic field
stems from the interplay  of the presence of orbital degeneracy  and 
anisotropy. We hope that this transition will be observed in further 
experiments on quasi one dimensional spin gapped materials
with orbital degeneracy.

{\it Note added}. After this work was completed, we became aware
of a work by Itoi $et$ $al$\cite{itoi} who also predict the extension of the
massless phase in the anisotropic region.

\newpage

\section{appendix}

In this Appendix, we shall compute the mass gap $M$ in the phase C, and obtain the 
asymptotics of $M$ in the vicinity of the critical surface $\Sigma$. As well known,
within perturbation theory the gap  defines  the scale
$t_M = \ln(1/Ma_0)$ where all the couplings blow up. Clearly,
 $t_M$ is given by the equation:
\be
X(t_M)= \infty.
\ee
Integrating (\ref{diffX}) and recalling the dynamic of $X$ in the phase C we find:
\be
t_M = \left( \int_{X^\star}^{\infty} + \epsilon \int_1^{X^\star}\right). 
\frac{dX}{X\sqrt{P(X)}}
\label{tMint}
\ee
where $\epsilon = \rm{sign}(s)$; $P(X)$ is given by Eq.(\ref{P}) and  has only two 
reals roots 0 and $X^\star\leq 1$. In the following $s$,$\mu$ and $G_3$ have to be
 understood as initial conditions.

Performing the integrals we obtain: 
\bea 
t_M &=& \sqrt{\frac{p}{|\mu|}}\left[ E(\alpha(0),k) 
- \epsilon E(\alpha(1),k)\right] \nonumber \\ 
&+&
\frac{1}{2} \sqrt{\frac{p}{|\mu|}}\left(\frac{u^{\star}}{p}-1\right)
\left[F(\alpha(0),k) 
- \epsilon F(\alpha(1),k)\right] \nonumber \\
&-&
\frac{G_3}{\mu}\left[\frac{1}{p+u^\star} -
 \frac{s}{2|G_3|(p+u^\star-1)} \right]
\label{tM}
\eea
where $F$ and $E$  the elliptic functions of the first  and
the second kind respectively, with parameters:
\bea
\alpha(u)&=&2\arctan\sqrt{\frac{u^\star-u}{p}} \nonumber \\
k&=&\sqrt{\frac{p+u^\star-{G_3^2}/2\mu {u^\star}^2}{2p}}
\label{parameters}
\eea
$u^\star$ and $p^2$ being given by:
\bea
u^\star&=&1/X^\star  \nonumber \\
p^2&=&{u^\star}^2-\frac{2G_3^2}{\mu u^\star}.  
\eea

Developping $t_M$ arround a point $(G_{1c},G_{2c},G_{3c})$ belonging to the 
critical surface $\Sigma$ between the phases B and C (see Fig. 2), we obtain 
the asymptotics
\bea
M &\sim& \Lambda \exp{(- C(\gamma_1)/\Delta^{2/3})} \hspace{1cm}
 \mbox{if}\;\;  G_1^c\!=\!G_2^c\!=\!0 \nonumber \\
M &\sim& \Lambda \exp{- C(\gamma_2)/\Delta}\hspace{1cm} 
\mbox{if}\;\; G_1^c\! \neq\! G_2^c
\label{gapasy}
\eea
where $\Delta$ is the Euclidian distance from $\Sigma$. The two constants
$C(\gamma_1)$ and $C(\gamma_2)$ are given by:
\bea
C(\gamma_1) &=& 0.6845 \ \ G_{3c}^{-1/3} (\cos\theta)^{-2/3}\nonumber \\
C(\gamma_2)  &=&
\left( \left(1+2{G_{3c}}^2/(G_{1c}-G_{2c})^2\right)  \cos^2\theta\right)^{-1/2} 
\label{c1c2}
\eea
where $\theta$ is the angle to the normal of $\Sigma$.

\end{document}